# Molecular dynamics simulation of the transformation of Fe-Co alloy by machine learning force field based on atomic cluster expansion

Yongle Li[1,*], Feng Xu[1], Long Hou[1], Luchao Sun[2], Haijun Su[3], Xi Li[1,4], Wei Ren[1,*]

[1] *Physics Department, State Key Laboratory of Advanced Special Steels, Materials Genome Institute, International Centre for Quantum and Molecular Structures, Shanghai University, Shanghai 200444, China*

[2] *Shenyang National Laboratory for Materials Science, Institute of Metal Research, Chinese Academy of Sciences, Shenyang 110016, China*

[3] *State Key Laboratory of Solidification Processing, Northwestern Polytechnical University, Xi'an 710072, China*

[4] *Shanghai Key Lab of Advanced High-temperature Materials and Precision Forming, Shanghai Jiao Tong University, Shanghai 200240, PR China*

*Emails: yongleli@shu.edu.cn; renwei@shu.edu.cn

**Abstract**

The force field describing the calculated interaction between atoms or molecules is the key to the accuracy of many molecular dynamics (MD) simulation results. Compared with traditional or semi-empirical force fields, machine learning force fields have the advantages of faster speed and higher precision. We have employed the method of atomic cluster expansion (ACE) combined with first-principles density functional theory (DFT) calculations for machine learning, and successfully obtained the force field of the binary Fe-Co alloy. Molecular dynamics simulations of Fe-Co alloy carried out using this ACE force field predicted the correct phase transition range of Fe-Co alloy.

**Key words:** Molecular dynamics, Atomic cluster expansion, Fe-Co Alloy, Density functional theory, Phase transition, Force field

**Introduction**

Alloy is usually a substance with metallic properties synthesized by two or more metal elements, or metal and non-metal elements through a specific method. According to the types of elements contained in the alloy, it can be divided into binary alloys, ternary alloys or multi-element alloys[1, 2]. The research on the phase transition of alloy materials has always been the focus of many scientific fields. Although numerous simulations of melting and solidification of alloys have been reported, progress in this area has been rather slow[3, 4].

Molecular dynamics (MD) is an effective means of simulating the phase transition of alloy materials, and molecular dynamics is widely applied in various fields such as physics, chemistry, biology and materials science[5-8]. At present, there are many methods for simulating the phase transformation of alloys by using molecular dynamics[9]. But the correctness of the molecular dynamics simulation results is largely limited by the precision of the chosen force field, which should benefit from the rapid development of empirical and semi-empirical many-body potentials describing metallic systems.

At present, the force fields that can be selected in the simulation of alloy materials include the Lennard-Jones (LJ) potential[10], embedded atom method (EAM) potential[11] and modified embedded atom method (MEAM) potential[12, 13], etc. Given the diversity and complexity of elements contained in alloys, the use of machine learning to fit force field parameters is faster, more promising, and hopefully more widely applicable than traditional force field development. Therefore, this paper uses the method of atomic cluster expansion (ACE)[14, 15] combined with first-principles density functional theory (DFT) calculations for machine learning to fit a force field that can be used for binary Fe-Co alloys. The melting and solidification mechanism of the binary Fe-Co alloy is revealed through the ACE force field, which expands the application of the ACE force field.

**Methods**

Atomic cluster expansion is a complete descriptor that can describe the local atomic environment of multicomponent materials. Some expressions for multivariate systems and non-orthogonal basis functions are given. Interatomic potentials with comparable precision to state-of-the-art machine learning potentials can be obtained by nonlinear functions from atomic clusters, which should converge to the precision of millielectron volt (meV). The principle of atomic cluster expansion is that the energy of the *i*-th atom is expressed as the coordinates related to other atoms[14].

$$E_i(\bm{r}_1,\bm{r}_2,\cdots,\bm{r}_N) = \sum_{v} c_v^{(1)} A_{iv} + \sum_{v_1 \geq v_2} c_{v_1 v_2}^{(2)} A_{iv_1} A_{iv_2} + \sum_{v_1 \geq v_2 \geq v_3} c_{v_1 v_2 v_3}^{(3)} A_{iv_1} A_{iv_2} A_{iv_3} + \cdots \quad (1)$$

$$A_{iv} = \sum_{j} \phi_v(\bm{r}_{ji}) \quad (2)$$

where $c$ is the expansion coefficient and $A_{iv}$ is the projection of the basis function on the atomic density.

Before fitting the ACE force field, we need to do preparatory work to create some sufficiently disordered systems, the atomic positions of which are random. For the selected systems, we obtain their energy, virial[16] and other information through first-principles density functional theory calculations. The DFT calculations were carried out using the Vienna Ab initio Simulation Package (VASP)[17], with projection-based augmented wave pseudopotentials[18], and Perdew-Burke-Ernzerhof (PBE)[19] generalized gradient approximation functional. In all calculations, the plane wave cutoff energy was chosen to be 400 eV, the atomic force convergence criterion was 0.01 eV·Å$^{-1}$, the energy convergence criterion was $10^{-6}$ eV, and the Brillouin zone was sampled with 13 × 13 × 13 k-points in a Monkhorst-Pack grid. After preparing various information of the system, we use the toolkit atomic simulation environment (ASE)[20] to integrate based on the python programming language.

The initial structure is obtained by constructing supercell of body-centered cubic (bcc) Fe-Co alloy with periodic boundary conditions. The ratio of Fe atoms to Co atoms is 1:1. As shown in Fig. 1, the established 10 × 10 × 10, 20 × 20 × 20, 30 × 30 × 30, and 37 × 37 × 37 Fe-Co alloy supercells of the volumes 28.4 × 28.4 × 28.4 Å$^3$, 56.8 ×

56.8 × 56.8 Å$^3$, 85.2 × 85.2 × 85.2 Å$^3$, and 105.08 × 105.08 × 105.08 Å$^3$ contain 2,000, 16,000, 54,000, and 101,306 atoms respectively. The MD simulation of 37 × 37 × 37 Fe-Co alloy supercell using the ACE force field is done with the Large-scale Atomic Molecular Massively Parallel Simulator (LAMMPS)[21]. The volume, density and other physical properties of this supercell at 300 K were simulated. In addition, four Fe-Co alloy supercells of different sizes were subjected to MD simulations under the isothermal-isobaric (NPT) ensemble. A Nosé-Hoover chain thermostat[22, 23] was used to control the system temperature and an Andersen method barostat[24] was used to control the pressure (1 atmosphere). The melting process of Fe-Co alloy supercells from 1 K to 2500 K is simulated with a time step of 1 fs and a heating rate of 0.5 K/ps[25]. The Velocity-Verlet algorithm[26] is used to solve the propagation classical equation of motion, and sample the real-time position and velocity of particles. Equilibrium simulations were performed at 1 K for 10 ps before the system was heated up. In addition, the solidification process of 37 × 37 × 37 Fe-Co alloy supercell cooling from 2500 K to 1000 K was simulated under the same conditions. Before the system cools down, equilibrium was simulated at 2500 K for 100 ps to ensure the system was in the liquid state. The Visual Molecular Dynamics (VMD)[27] package and the Open Visualization Tool (OVITO) package were used for structure monitoring and trajectory analysis of various output data generated in MD simulations.

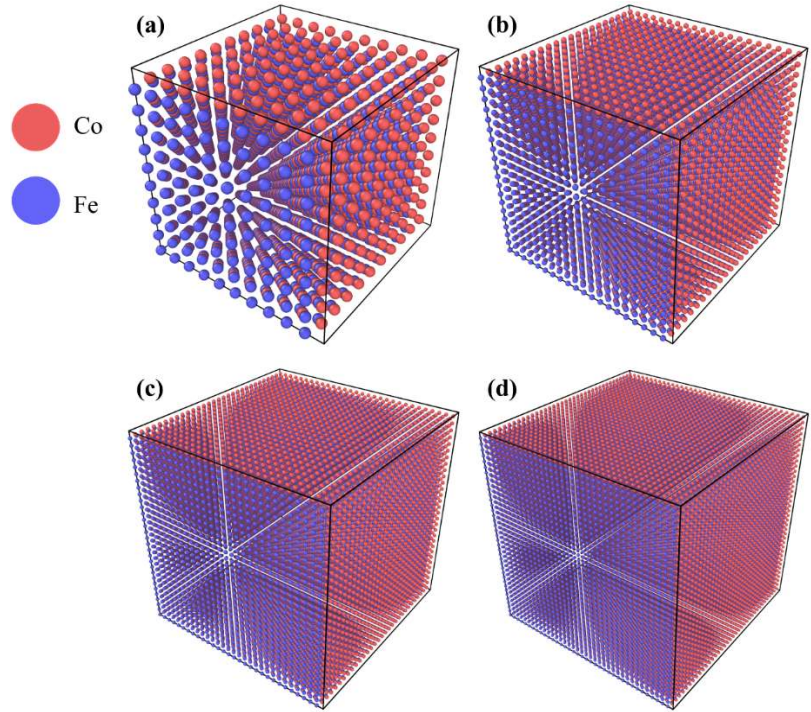

Fig. 1. (a) supercell structure of Fe-Co alloy 10 × 10 × 10, (b) supercell structure of Fe-Co alloy 20 × 21 × 20, (c) supercell structure of Fe-Co alloy 30 × 30 × 30 and (d) supercell structure of Fe-Co alloy 37 × 37 × 37.

## Results and discussion

### Verification of ACE force field at 300 K

In order to verify the correctness of the ACE force field, we simulated various properties of the 37 × 37 × 37 Fe-Co alloy supercell at 300 K, including volume, density and energy. As shown in Fig. 2, the 37 × 37 × 37 Fe-Co alloy is heated starting at ultralow temperature 1 K to 300 K at 0.5 K/ps and equilibrated at 300 K for 10 ns after a period of equilibrium. The insets zoom in the results when the temperature is goes up. In the simulation of equilibrium at 300 K for 10 ns, its fluctuation is found to be reasonably small. At this time, the temperature, volume, density and energy of the 37 × 37 × 37 Fe-Co alloy supercell are recorded in Table 1, which are in excellent agreement with the experimental values[28].

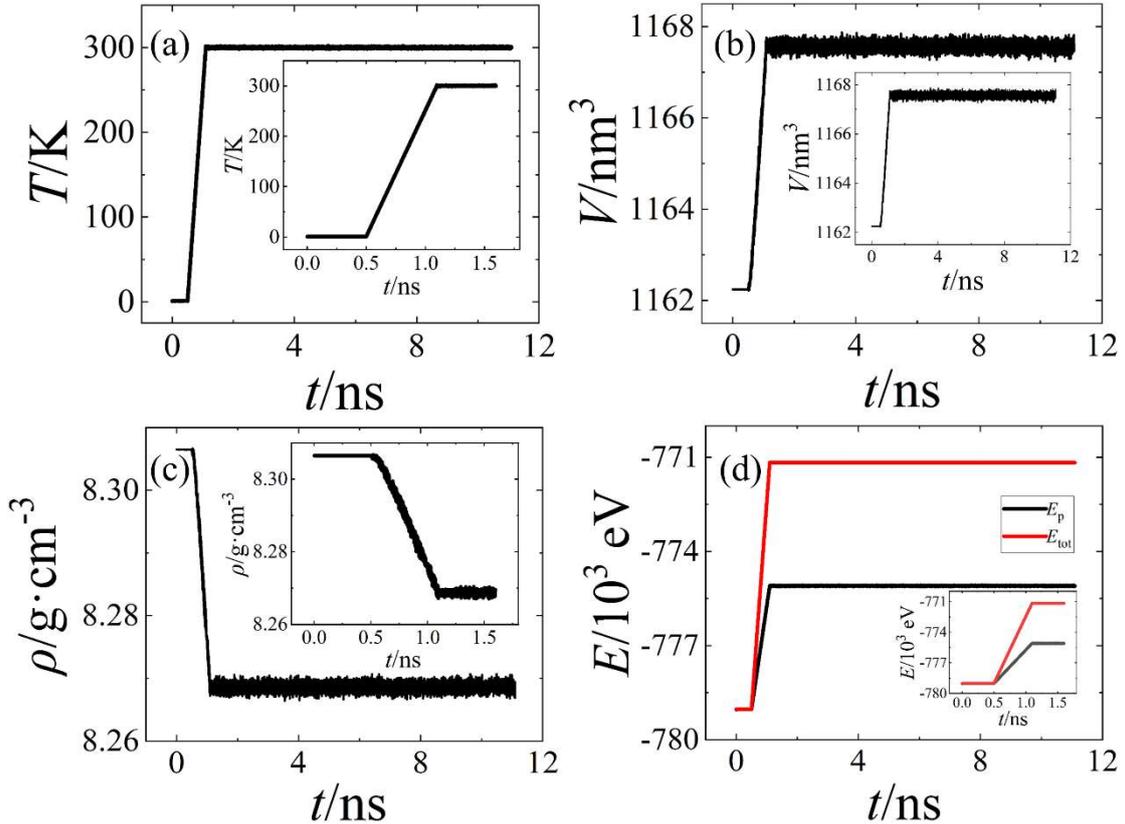

Fig 2. (a) Temperature; (b) volume; (c) density and (d) energy change curves with evolution time of 37 × 37 × 37 Fe-Co alloy supercell structure using ACE force field for molecular dynamics. The insets are the enlarged pictures of the heating stage.

Table 1. Parameter values and fluctuations of 37 × 37 × 37 Fe-Co alloy supercell from molecular dynamics equilibrium under ACE force field and experimental values

of related parameters.

| Property | This Work | Fluctuation | Experimental [28] |
|---|---|---|---|
| Temperature (K) | 300 | ±2 | |
| Volume (nm$^3$) | 1167.56 | ±0.31 | 1172.57 |
| Density (g/cm$^3$) | 8.269 | ±0.002 | 7.9 - 8.0 |
| Potential Energy (eV) | -775097 | ±26 | |
| Total Energy (eV) | -771168 | ±4 | |

We have also used the ACE force field to simulate the bulk elastic modulus of Fe-Co alloy at 300 K, and searched for the structure with the lowest unit cell energy of Fe-Co alloy by changing the lattice parameters. Due to the relatively small number of atoms in the unit cell, there may be size-induced errors in the simulation results, so we expanded the Fe-Co alloy cell to 37 × 37 × 37. The Fe-Co alloy equilibrium lattice parameter (at the minimal point of Fig. 3) is denoted by $a_0$, and the potential energy of the unit cell is denoted by $E_p$. The bulk modulus[29] is defined as follows:

$$B \equiv -\frac{dP}{dV/V} \tag{3}$$

$V$ and $P$ correspond to the volume and pressure, respectively. And for a cubic unit cell we have:

$$P = -\frac{d\varepsilon}{dV} = -\frac{M}{3a^2}\frac{dE}{da} \tag{4}$$

At this condition:

$$B = \frac{M}{9a_0}\frac{d^2E}{da^2}\bigg|_{a_0} \tag{5}$$

where $M$ in the formula is the number of atoms in the cubic unit cell.

As shown in Fig 3, we simulated the variation of the unit cell potential energy of the Fe-Co alloy upon changing the lattice parameter. At the minimum, the equilibrium lattice parameter $a_0$ = 2.842 Å, and the unit cell potential energy $E_p$ = -15.38 eV. We performed a fifth-order nonlinear fitting on the potential energy of the unit cell corresponding to different lattice parameters, and obtained the expression of potential energy and lattice parameters as: $E = b_0 + b_1a^1 + b_2a^2 + b_3a^3 + b_4a^4 + b_5a^5$. From formula (5), we obtained a bulk modulus of 2.376 eV/Å$^3$ for the Fe-Co alloy which is close to the experimental value[30].

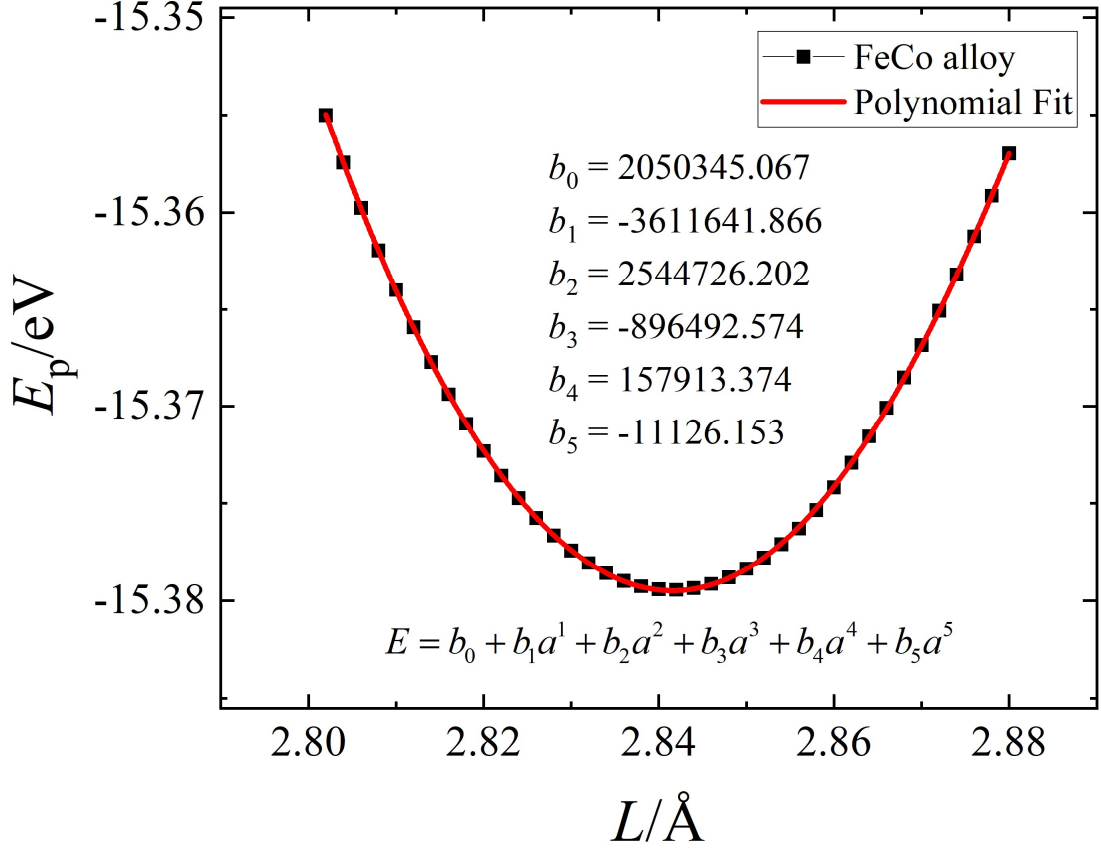

Fig. 3. The variation curve of unit cell potential energy of Fe-Co alloy with lattice parameters.

**Phase transitions - Melting and Freezing points**

We simulated the heating up of 10 × 10 × 10, 20 × 20 × 20, 30 × 30 × 30 and 37 × 37 × 37 Fe-Co alloy supercells from 1 K to 2500 K, respectively. As shown in Fig. 4, the melting point range of the system can be judged by observing the volume change with temperature. And we found that with the increase of the number of atoms in the system, the fluctuation of the melting point becomes smaller and smaller. Taken the 37 × 37 × 37 Fe-Co alloy supercell as an example, its melting point is about 1766 K, which agrees very well with the experimental values and the results of other force field simulations. The melting points and their fluctuations for Fe-Co alloy supercells of different sizes, as well as the experimental values and the results simulated by other force fields are listed in Table 2.

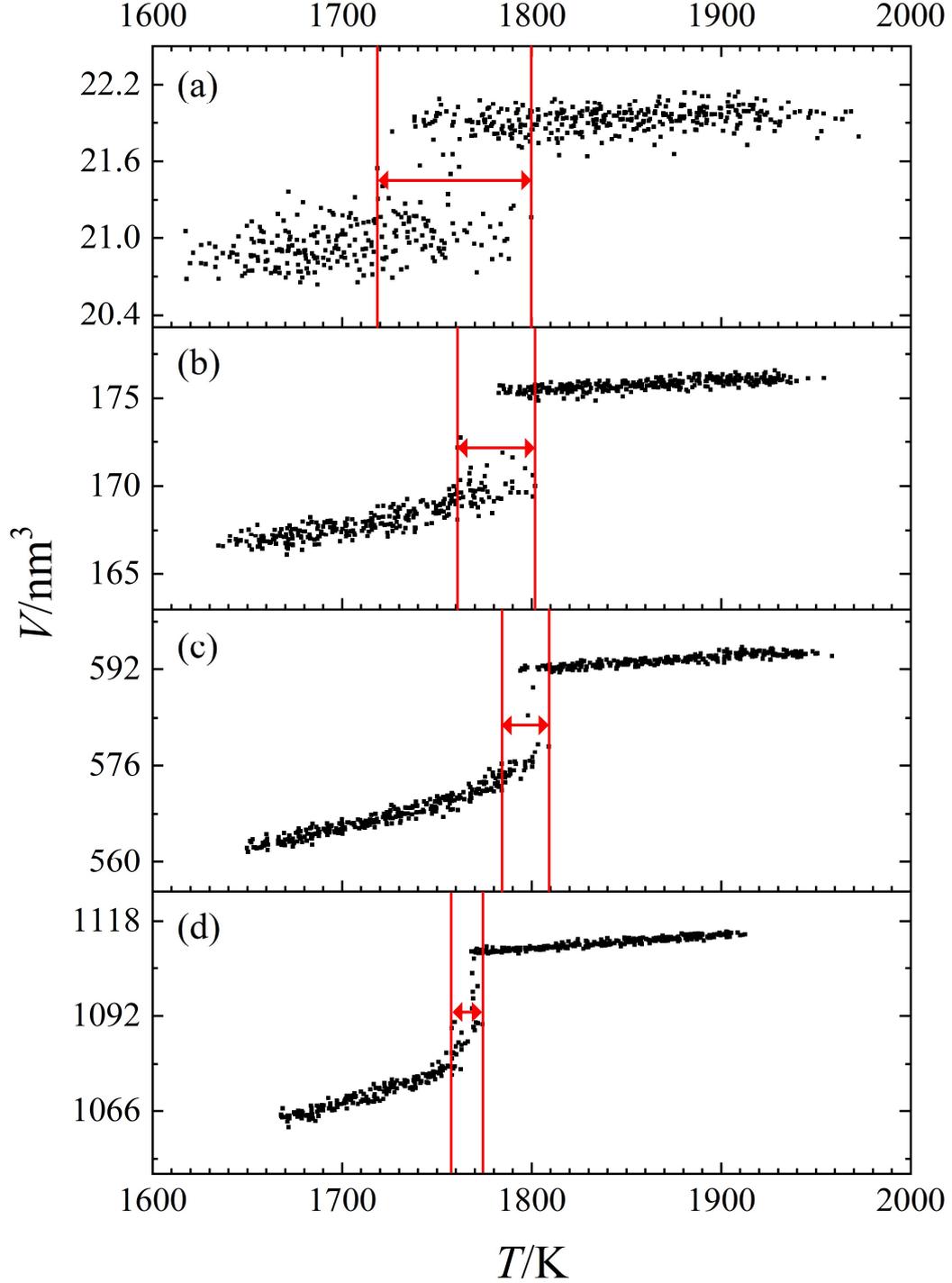

Fig. 4. (a) to (d) are the change in volume with temperature near the melting point during warming up of 10 × 10 × 10, 20 × 20 × 20, 30 × 30 × 30 and 37 × 37 × 37 Fe-Co alloy supercells. The red lines indicate the phase transition interval.

Table 2. Melting points and fluctuations of Fe-Co alloy supercells with different sizes using ACE force field, as well as experimental values and other force field simulation result.

| Size of the system | Melting point (K) | Fluctuation (K) |
| --- | --- | --- |
| This Work_10 × 10 × 10 | 1763 | ±37 |
| This Work _20 × 20 × 20 | 1781 | ±20 |
| This Work _30 × 30 × 30 | 1796 | ±13 |
| This Work _37 × 37 × 37 | 1766 | ±8 |
| Experimental[31] | 1727,1724 | |
| Experimental[32] | 1750 | |
| MEAM[33] | 1700 | |

In order to understand the phase transformation mechanism of Fe-Co alloy in depth, we present the radial distribution function (RDF) $g(r)$[34-36] of different temperature for 37 × 37 × 37 Fe-Co alloy supercell during the heating process. The first peak in RDF represents the average number of atoms at the nearest neighbors of the target atom, and the second peak represents the average number of atoms at the next-nearest neighbors. When the Fe-Co alloy is melted, as in a liquid state, such phase only has short-range order. Therefore, the RDF of Fe-Co alloy after melting appears featureless after the nearest neighbor, and the peaks are wide and smooth. As the thermal motion of atoms becomes more and more intense with the increase of temperature, the heights of wave crests become lower and lower, while the wave peaks become wider and wider. In Fe-Co alloy with a bcc structure with a ratio of Fe:Co atomic numbers of 1:1, the nearest neighbor atoms of the body-centered atoms are located at the 8 vertices of the cubic unit cell, and the second nearest neighbor is the center of the adjacent unit cell. As shown in Fig 5, during the heating process, we found that as the temperature increased, the first peak began to decrease slowly, and the second split peak gradually became smoother, which means that the system is undergoing a phase transition.

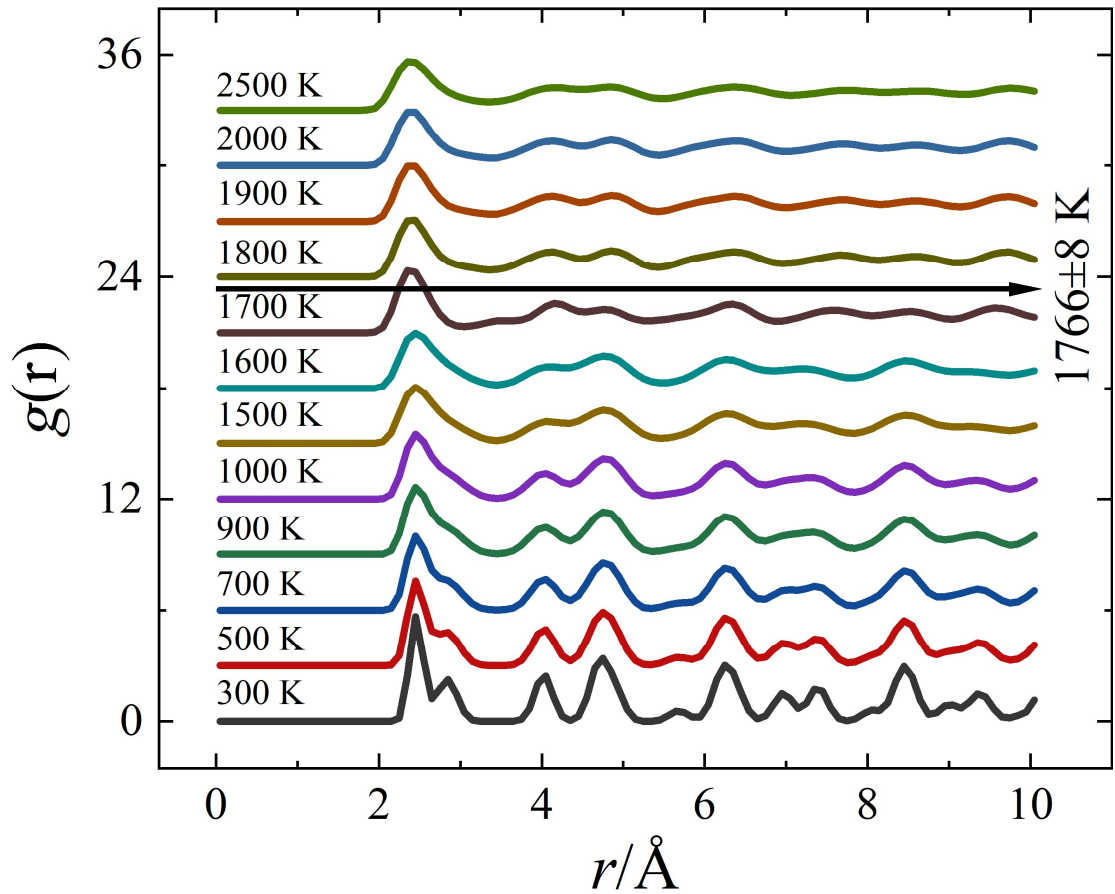

Fig. 5. Radial distribution function of Fe-Co alloy 37 × 37 × 37 supercell at different temperatures.

As shown in Fig 6(a), we also used the ACE force field to simulate the cooling process (same rate as heating 0.5 K/ps) of the 37 × 37 × 37 Fe-Co alloy supercell from 2500 K to 1200 K. Similarly, the freezing point of Fe-Co alloy is about 1665 K and the fluctuation is 5 K by observing the volume change with temperature. From Fig 6, we found that the volume, density and potential energy of the Fe-Co alloy near the phase transition point all behave to form closed loop of hysteresis.

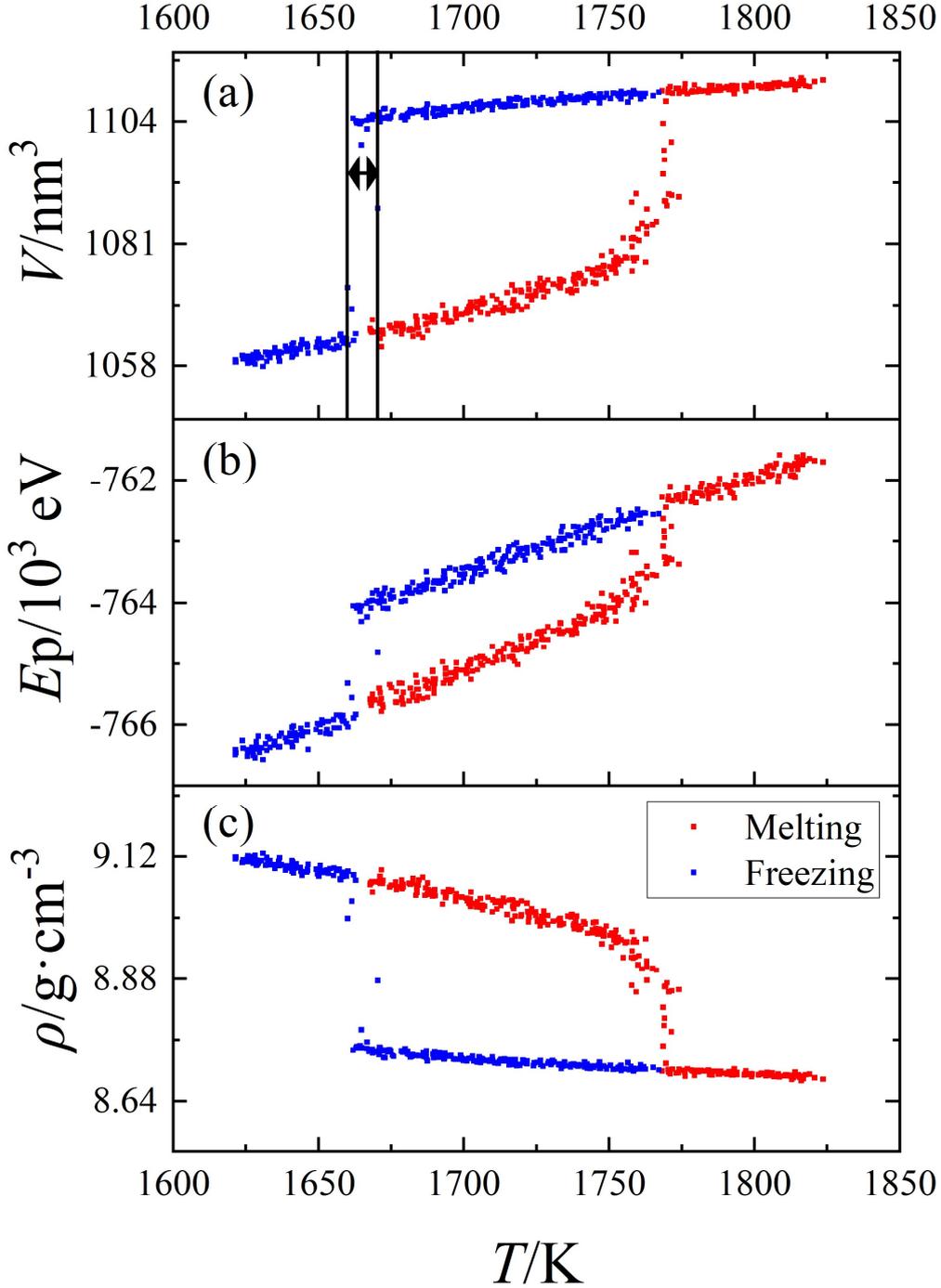

Fig. 6. Hysteresis loops of the (a) volume, (b) density and (c) potential energy of the 37 × 37 × 37 Fe-Co alloy supercell versus temperature. The black lines indicate the phase transition interval.

As shown in Table 3, we found that the solidification and melting of metals or alloys are actually asymmetrical[4], and the freezing point is smaller than the melting point. According to the classical nucleation theory[37, 38], this is because the liquid system will be subcooled[39, 40] during the solidification process and the interface where the new phase appears will prevent the solidification from nucleation, and conversely, the equilibrium of surface energy at the interface between the phase and the phase during the melting process will prevent the solid from melting, so the melting and freezing points are not symmetrical and slightly different in temperature.

Table 3. Melting point and freezing point of 37 × 37 × 37 Fe-Co alloy supercell and their fluctuations

| Properties | Temperature (K) | Fluctuation (K) |
| --- | --- | --- |
| Melting point | 1766 | ±8 |
| Freezing point | 1665 | ±5 |

**Conclusion**

In this work, we have successfully fitted the force field of the binary Fe-Co alloy by machine learning combined with atomic cluster expansion and the first-principles density functional theory. The properties of Fe-Co alloy at 300 K were simulated by using the ACE force field, and the correctness of the obtained parameters was verified. From the ACE force field molecular dynamics, the melting point of Fe-Co alloy is 1766 ± 8 K, and the freezing point is 1665 ± 5 K. The hysteresis of the melting and freezing points of the Fe-Co alloy was found, which is consistent with the classical nucleation theory. The ACE force field is easier to fit than the traditional empirical and semi-empirical multi-body force fields, which may widely expand its application.

**Acknowledgements**

This work was supported by the National Natural Science Foundation of China (No. 22173057, 52130204, 12074241, and 11929401), Science and Technology Commission of Shanghai Municipality (Grants No. 21JC1402700, No. 20501130600, No. 20QA1401000, No. 21JC1402600, and No. 22XD1400900), High Performance Computing Center, Shanghai University, and Key Research Project of Zhejiang Laboratory (Grant No. 2021PE0AC02).